# Optical Property Study of Charge Compensated (Si, Na) Co-doped ZnO

Tulika Srivastava[1], Gaurav Bajpai[1], Gyanendra Rathore[1], Prashant Mishra[3], Shun Wei Liu[2], Sajal Biring[2]*, Somaditya Sen[1,2,3]**

[1]Metallurgy Engineering and Materials Science, Indian Institute of Technology Indore, Khandwa Road, Simrol, Indore, India

[2]Electronic Engineering Department and Organic Electronic Research Centre, Ming Chi University of Technology, New Taipei City, Taiwan

[3]Department of Physics, Indian Institute of Technology Indore, Khandwa Road, Simrol, Indore, India

**Email ids:** *sens@iiti.ac.in, biring@mcut.edu

**Abstract:**

ZnO is co-doped with $Na^+$ and $Si^{4+}$ in the ratio 2:1. The ratio was intentionally chosen so that net valence state of dopant theoretically matches that of host. This is to avoid dependence in the amount of oxygen vacancies/interstitials arising out of cationic valence state of the dopant. With such a combination, modifications in structural and optical properties do not depend on excess or deficit of the dopant's charge state. For lower doping, $Na^+$ ions behave as interstitial sites which enhance strain, lattice disorder and thereby creating defects. Formation of interstitial defects leads to reduction in bandgap energy and produce orange-red luminescence. For higher doping, $Na^+$ starts substituting at $Zn^{2+}$ site which helps in reducing strain and lattice disorder and thereby increases bandgap. Inspite of presence of $Si^{4+}$ with higher charge, there is a gradual increase in oxygen vacancies due to lattice disorder.

**Introduction:**

Zinc oxide (ZnO) has received much attention due to its wide band gap (3.37 eV) and high excitonic energy (60 meV). It is non-toxic, economic and environmental friendly [1]. It has applications in various areas like room-temperature ultraviolet (UV) lasers [2], LEDs [3], solar cells [4], light detectors [5], field-effect transistors (FETs) [6], photo-catalysis [7] etc. It is a semiconductor having good chemical stability against hydrogen plasma [8]. It is suitable for photovoltaic applications due to its high-electrical conductivity and optical transmittance in visible region of solar spectrum, which is primarily important in solar cell fabrications. Doping offers a method to tailor electrical, optical and magnetic properties of ZnO [9]. As an extrinsic semiconductor, n-type conductivity is common in ZnO [10]. For achieving n-type conductivity in ZnO, group III elements (B, Al, Ga and In), group IV elements (Si, Ge, Sn) on Zn- site and group VII elements (F,Cl) on O-site has been reported [11–15]. But, p-type conductivity in ZnO is difficult to achieve [10]. However, doping group I elements (Li, Na, K, Cu, Ag) on Zn-site can result in p-type nature [16,17]. Similarly Zn vacancies and group V elements (N, Sb and As) at O-site may also produce p-type conductivity in ZnO [18–20].

Three factors, i.e., dopant formation energy, dopant ionization energy and dopant solubility determine properties of modified ZnO [21]. Doping modifies native defects or introduces new defect states, thereby affecting luminescence of ZnO. Red emission in Co doped ZnO [22], yellow-orange-red emission in Mn doped ZnO [23], blue emission in Cu doped ZnO[24], green emission in Tb doped ZnO [25], red [26,27] and blue emission in Ce doped ZnO [28] etc. have been reported. Bandgap decreases by doping Cr, Ni, Cd but increases with Mg, Co and Mn doping [29].

In a previous work it was reported that $Si^{4+}$ doped ZnO leads to enhancement in bandgap and reduction in oxygen vacancies due to higher charge of $Si^{4+}$ in comparison to $Zn^{2+}$ [30]. In this manuscript, $Na^+$ is co-doped in $Si^{4+}$ doped ZnO to see effect of compensation of extra charge due to Si. For every $Si^{4+}$ two $Na^{1+}$ are needed. Hence, a ratio of Na:Si= 2:1, was chosen intentionally to achieve this goal. Therefore, $Zn_{1-x}Na_{2x/3}Si_{x/3}O$ was synthesized using sol-gel process. Defect structure and optical properties of these material has been studied in detail.

**Experimental:**

$Zn_{1-x}Na_{2/3}Si_{1/3}O$ with x=0 (Z0), 0.013 (ZSN1), 0.020 (ZSN2) and 0.027 (ZSN2) respectively have been synthesized by standard Pechini sol-gel process followed by solid state sintering. ZnO powder is dissolved in $HNO_3$ (Alfa Aesar, purity 99.9%) and an appropriate amount of orthosilicate (($C_2H_5O)_4Si$) and Sodium nitrate ($NaNO_3$) is added. A homogeneous solution was prepared by prolonged stirring. A polymeric solution was formed by heating 1 g of citric acid (Alfa Aesar, 98%) and 1 mL of 20% concentrated glycerol (Alfa Aesar, 99.9%) in de-ionized water at 70 °C for 4 h. These Zn/Si/Na solutions were then added to polymeric solutions. Resultant solutions were stirred and heated on hot plates at ~60 °C. To polymeric chains, Zn/Si/Na ions homogenously got attached. Solutions gradually dehydrated to form gels in ~4hr, which were burnt on hotplates in ambient conditions to form dark powders. Resultant powders were decarbonized and denitrified by heating in air at 450 °C for 6 h and then annealing at 600°C for 2 h.

X-ray diffraction studies were carried out using a Bruker D2-Phaser diffractometer. Agilent UV–vis spectrometer (model Carry 60) was used to analyse electronic bandgap. Room temperature photoluminescence measurement was investigated using Dongwoo Optron DM 500i.

**Result and Discussion:**

A dominant hexagonal wurtzite ZnO structure was revealed from XRD studies. Some minor reflections coming from zinc blend was also observed (Figure 1 (a)). No secondary phases were found related to simple or complex oxides of Zn, Si and Na. However, for higher substitution (x ≥0.027) secondary phases appear. Average crystallite size was calculated using Scherer formula (inset of figure 1(a)). Field Emission Scanning Electron Microscopy (FESEM) images show agglomerated particles for all samples. Particle size (average) was estimated using Image J software. Average particle sizes first decrease from 0.7 to 0.148μm but thereafter increase to 0.3 μm (Figure 1(b)). This trend is consistent with crystallite size obtained from XRD data. However, XRD values are much smaller than values obtained from SEM studies which reveal larger agglomerated particles composed of smaller crystallites.

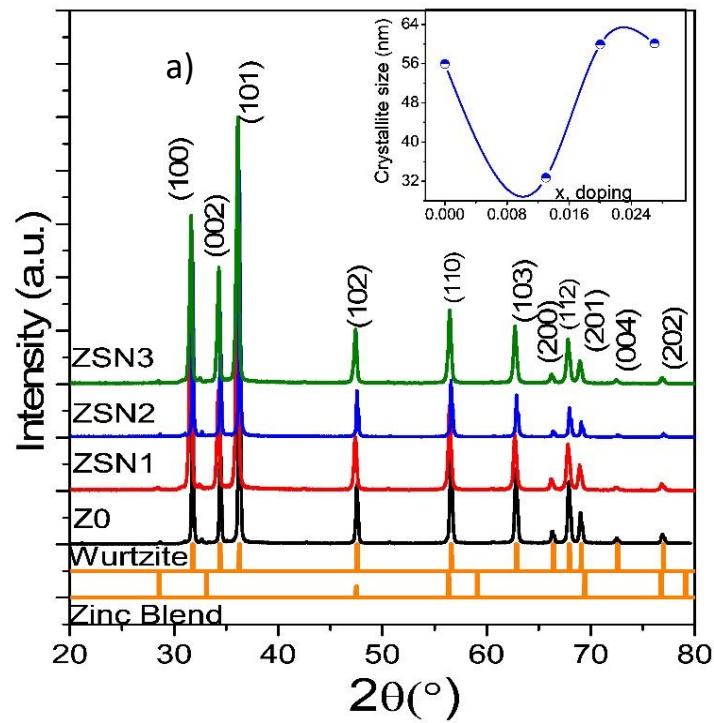

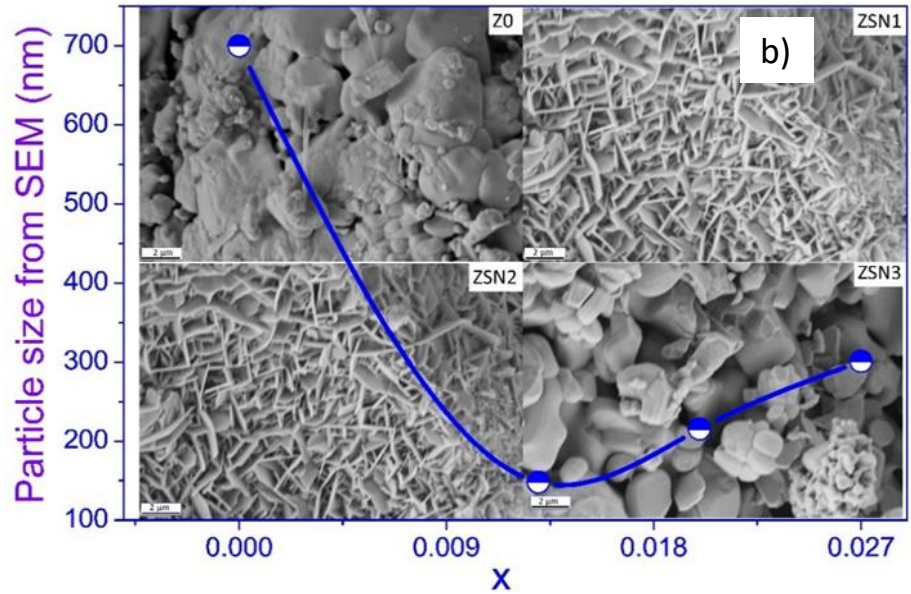

*Figure 1: a) XRD spectrum of Z0, ZSN1, ZSN2 and ZSN3 (inset is crystallite size) b) FESEM image of Z0, ZSN1, ZSN2 and ZSN3 (inset is variation of particle size with doping)*

Rietveld refinement was done using GSAS software to estimate changes in lattice parameter, strain and crystallite size with silicon substitution. Pseudo-Voigt profile was chosen for peak shape. Background was modelled using a 6-coefficient polynomial function. For lower doping (ZNS1), strain increases but lattice parameters decrease (figure 2(a, b)). However, for higher doping, strain reduces but lattice parameters increase. Increased strain inhibits particle growth and reduces particle as well as crystallite size. Reverse phenomenon happens for higher doping.

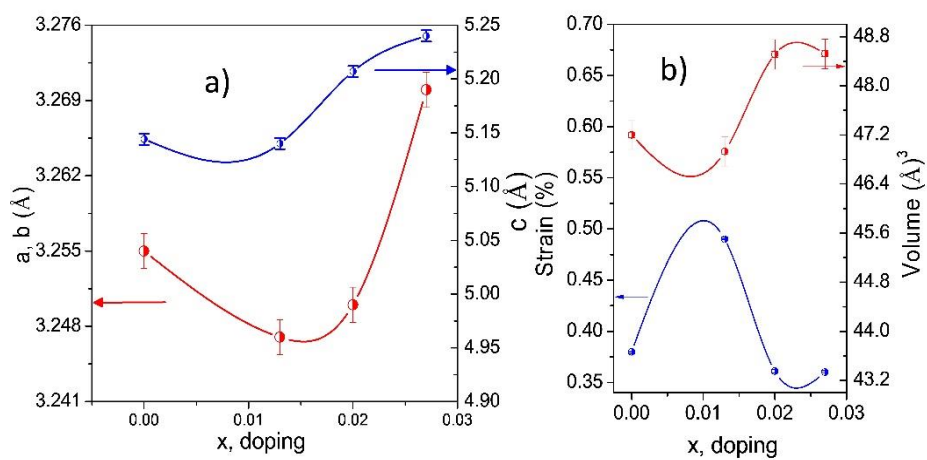

*Figure 2: Variation of (a) lattice parameter (b) Volume and strain with $Na^{1+}/Si^{4+}$ co-doping.*

Reports [31,32] on Na doped ZnO suggested that Na prefers to occupy interstitial sites whereas higher doping leads to increase in substitutional site. Similar trend is observed in

$Si^{4+}/Na^+$ doped ZnO samples. With minimal $Si^{4+}/Na^+$ doping (ZSN1), $Na^+$ prefers to occupy interstitial sites which pressurizes lattice, reduces lattice parameters and thereby enhances lattice strain of system. For higher doping, substitution increase due to proper rearrangement and coupling of Na and Si in lattice, reducing strain. Due to higher ionic radii of Na, volume as well as lattice parameter increases.

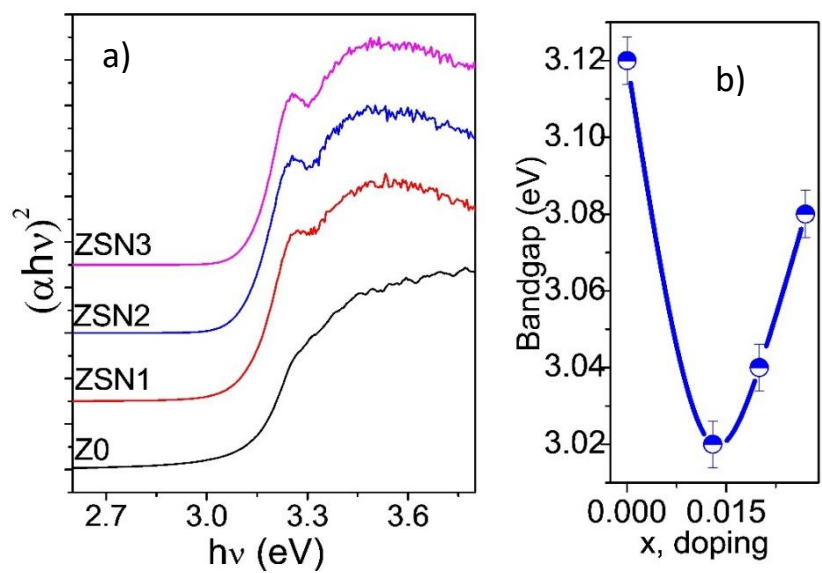

*Figure 3: (a) Absorption spectra of Z0, ZSN1, ZSN2 and ZSN3 (b) Variation of bandgap with $Na^+/Si^{4+}$ co-doping.*

Optical bandgap properties were studied using UV-Vis spectroscopy (figure 3(a)). Absorption coefficient, α, is related to bandgap of a material. As ZnO is a direct bandgap material, extrapolation of linear part of $(\alpha h\nu)^2$ vs $h\nu$ graphs were plotted until it meets x-axis gives value of bandgap. Band gaps were found to be decreasing from 3.12 eV in Z0 to 3.11 eV in ZSN1. Thereafter, bandgap increases for higher doping (figure 3 (b)).

To understand effect of structural disorder on bandgap, Urbach energy ($E_U$) has been calculated. A plot of $\ln(\alpha)$ vs $h\nu$ was plotted and a linear fitting was done on linear part of curve (figure 4(a)). Urbach energy ($E_U$) increases from 83 meV in Z0 to 125 meV in ZSN1 and then decreases to 60 meV for ZSN3. These changes in Urbach energy ($E_u$) (structural disorder) effectively modify bandgaps.

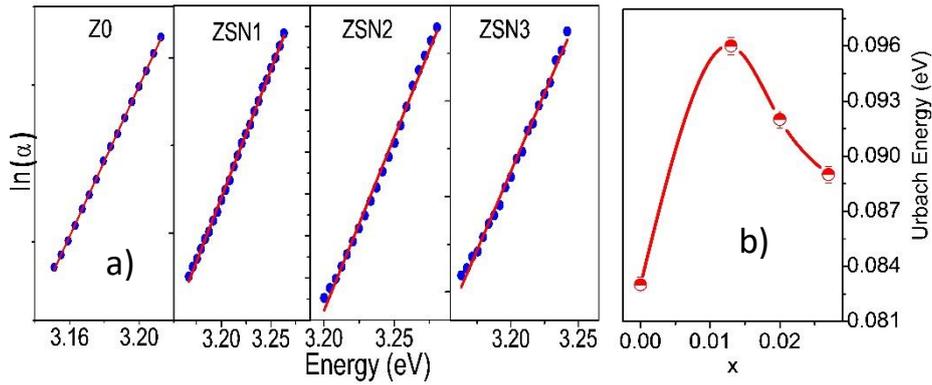

*Figure 4: (a) Linear fitting to estimate Urbach energy (b) Variation of Urbach energy with co-doping.*

$E_U$ values change inversely with optical band gap ($E_g$) (figure 4 (b)). Variation in bandgap might be due to changes in Urbach tail width. Increase/decrease of $E_U$ suggests enhancement/reduction in atomic structural disorder of ZnO, related to decrease/increase of optical bandgap respectively with increasing $Na^+/Si^{4+}$ concentration. Changes in lattice disorder are also confirmed by strain variation with $Na^+/Si^{4+}$ co-doping as calculated from XRD. Variations in lattice irregularity and strain hints at modification of defects state which are later confirmed from photoluminescence studies.

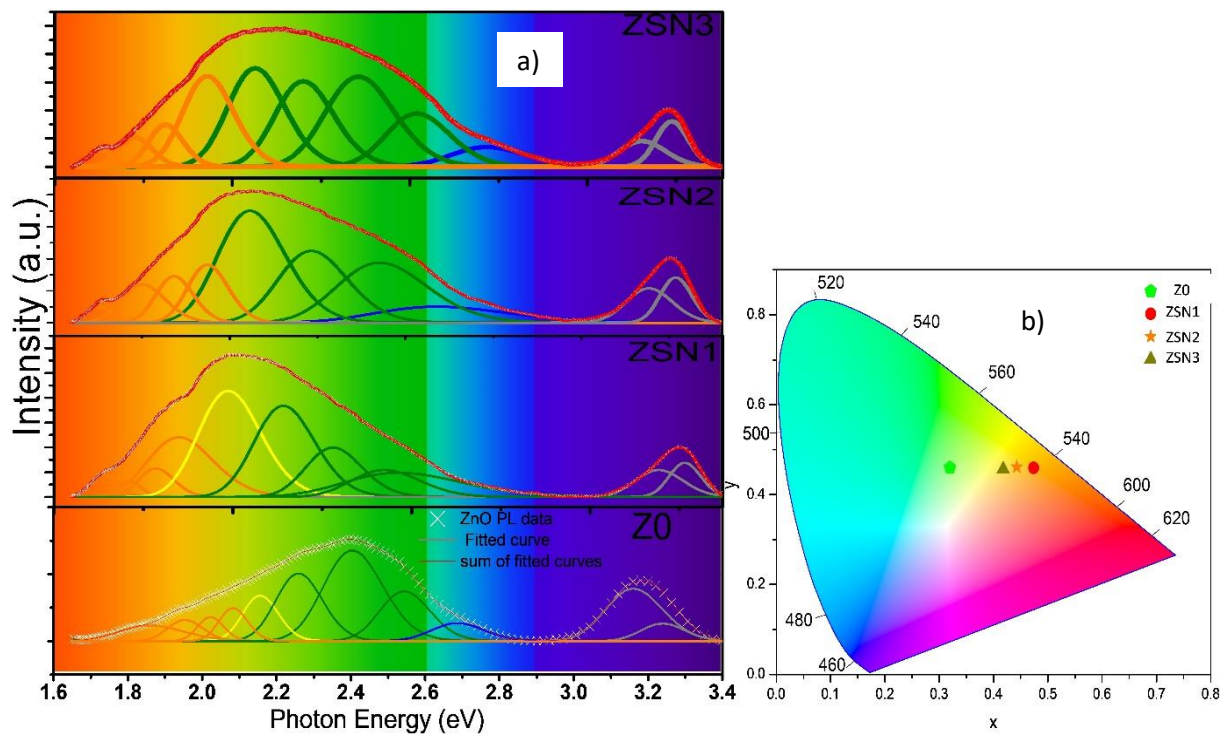

*Figure 5: (a) Gaussian fitting of PL spectra Z0, ZSN1, ZSN2 and ZSN3 (b) Chromaticity diagram.*

Room temperature PL of samples was studied in order to understand effect of $Na^+/Si^{4+}$ co-doping on optical properties of modified ZnO (figure 5 (a)). A relatively sharper, near band edge (NBE) centred at ~390 nm in comparison to a wider and broader, deep level emission (DLE) extending laterally from 450 nm to 750 nm, is observed in all samples. Note that DLE of Z0 is centred at 518 nm whereas other doped samples are centred at ~600 nm: a clear red shift in substituted samples is observed. To further confirm this, chromaticity coordinates have been calculated and plotted on 1931 standard chromaticity diagram.

Calculated coordinates are (0.32, 0.46), (0.47, 0.45), (0.44, 0.46), and (0.41, 0.45) for Z0, ZSN1, ZSN2, and ZSN3, respectively which indicates an enhancement of orange-red emission (figure 5 (b)). It confirms shifting of color emission from green region to orange-red-yellow region with $Na^+/Si^{4+}$ co-doping.

DLE is a broad spectrum. It is composed of multiple peaks corresponding to UV (>3.1 eV), violet (3–3.1 eV), blue (2.50–2.75 eV), green (2.17–2.50 eV), yellow (2.10–2.17 eV) and orange-red (<2.1 eV) colour ranges. Hence to quantify specific defect states, broad DLE spectrum as well as NBE was de-convoluted into several peaks belonging to the above colour ranges using Gaussian fittings. Colour of light emitted is a resultant of mixture of such colours.

NBE spectra of all samples are deconvoluted into two peaks: Z0 (3.16 eV & 3.24 eV); ZSN1 (3.18 eV & 3.26 eV), ZSN2 (3.18 eV & 3.27 eV) and ZSN3 (3.16 eV & 3.24 eV) respectively. First peak is due to donor acceptor pair while second peak is due to free excitons.

DLE of Z0 is deconvoluted into 9 peaks located at 2.69 eV (blue emission), 2.25 eV, 2.4 eV, & 2.54 eV (Green emission), 2.08 eV & 2.15 eV (yellow emission) and 1.85 eV, 1.95 eV & 2.02 eV (orange-red emission). Similarly, for substituted samples each color range is composed of a set of peaks contributing to same:

Blue: (ZSN2: 2.63 ZSN3: 2.71 eV)

Green (ZSN1: 2.28, 2.42, 2.48; ZSN2: 2.42, 2.23; ZSN3: 2.5, 2.35, 2.20)

Yellow (ZSN1: 2.14 eV, 2.15 eV & 2.15 eV)

Orange-red (ZSN1: 1.64, 1.70, 1.78, 1.84 & 1.98; ZSN2: 1.63 eV, 1.74 eV, 1.93, 1.84 & 2.06 eV; ZSN3: 1.98 eV, 2.05 eV; ZS2: 1.75 eV, 1.77 eV, 1.82 eV, 1.89 eV, 1.98 eV, 2.05 eV; ZS3: 1.81 eV, 1.84 eV, 1.90 eV, 1.96 eV, 2.02 eV, 2.09 eV)

Marked decrease in intensity ratio of $I_{NBE}/I_{DLE}$ from Z0 (0.6) to ZSN1 (0.35) is observed. Thereafter, an increase for ZSN2 and ZSN3 signify enhancement in defect states with minimal co-doping of Na/Si and followed by a reduction. $P_{NBE}/P_{DLE}$ was calculated (where $P_{NBE}$ is total area of peaks in NBE region and $P_{DLE}$ is total area of peaks in DLE region) and was found to follow same trend as $I_{NBE}/I_{DLE}$. These trends exactly follow strain (calculated in structural studies) and Urbach energy. This indicates that defect states created by Na in ZSN1 is responsible for strain and lattice disorder. For higher doping ZSN2 & ZSN3, silicon dominates and reduces overall defects.

To understand contribution of each colour emission to total spectrum, area of each peak was added and grouped into $A_{UV}$, $A_B$, $A_G$, $A_Y$, and $A_{O–R}$ for UV, blue, green, yellow, and orange-red. Colour fraction $P_N = P_{colour}/P_{UV}$ of each colour was calculated using already published procedure [30].

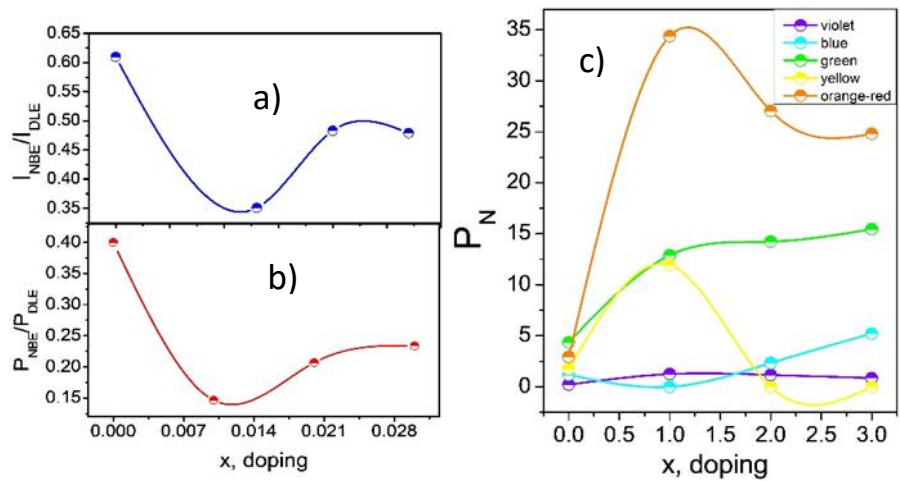

*Figure 6: Variation of (a) $I_{NBE}/I_{DLE}$ (b) $P_{NBE}/P_{DLE}$ with x (c) $P_N$ Vs x*

Green emission steadily increases from undoped samples to highly doped samples. Oxygen vacancy increases with $Na^+$ co-doping which signifies that charge compensation is incapable of reducing oxygen vacancies. Orange-red emission enhanced significantly for lower doping and thereafter decreased for higher doping. This trend clearly follows strain calculated from XRD. For lower doping, Na enters into interstitial sites ($Na_i^{\bullet}$) and increases strain in system. Structural disorder might have produced oxygen interstitials which enhances orange-red emission. For higher doping, with increased amount of Si more lattice space is created due to extremely small size of Si. Na and Si rearrange positions to manage space. This reduces $O_i$ defects in lattice. Yellow emission, another signature of $O_i$, follows similar trend as orange-red emission. This further confirms our proposed mechanism. Violet and blue emission is approximately invariant compared to other colors. This indicates minimal presence of $Zn_i$ and $V_{Zn}$ defects.

**Conclusion:**

Co-doping of $Na^+/Si^{4+}$ in ZnO at ratio 2:1 is done to achieve charge difference compensation between dopant and host ion. At lower doping (ZSN1), Na enters into interstitial sites which increase strain and Urbach energy thereby reducing bandgap of material. Increase in interstitial defects is also confirmed by enhancement in orange-red emission in photoluminescence spectra. For higher doping (ZSN2 and ZSN3), $Na^{1+}$ and $Si^{4+}$ substitute $Zn^{2+}$ which reduces strain and Urbach energy thereby increasing bandgap of material. Further reduction in orange-red emission indicates interstitial defects reduction. Inspite of charge difference compensation and presence of higher charge dopant ($Si^{4+}$), there is enhancement in oxygen vacancies as confirmed by increase in green emission.